\documentclass{ws-p8-50x6-00}

\def\be{\begin{equation}}
\def\ee{\end{equation}}
\def\bea{\begin{eqnarray}}
\def\eea{\end{eqnarray}}

\def \e{{\rm e}}

\begin{document}

\title{HARD INELASTIC INTERACTIONS AT PARTON AND HADRON LEVEL} 
\author{G. Calucci, E. Cattaruzza}

\address{Dipartimento di Fisica teorica, Universit\`a di Trieste, Strada
Costiera 11. I-34014 and I.N.F.N. Sezione di Trieste, Italy } 


\maketitle\abstracts{
 In the study of multiple scattering of partons in hadron-hadron collisions the
 possibility of a hard inelastic process at the parton level is included in its
 simplest possible way, $i.e.$ including the $2 \to 3$ transition.
 The specific physical process to which the treatment is applied is the 
inelastic
 collision of a nucleon with a heavy nucleus}

\section{ Motivations}
\vskip .5pc

 As a part of a wider investigation on the contribution of multiple hard 
 scattering OF partons in hadron-hadron
 collisions the production of parton already at hard, perturbative level
 is considered.
 A definite part of the problem is investigated: the elementary parton collision
 can give rise only to one more parton, the production and reabsorption are
 taken into account, the process has presumably its largest relevance in
 nucleus-nucleus collisions, but for simplicity only the hadron-nucleus
 collision is investigated.
 For the moment, what is shown is the
 possibility of studying the problem with the re-interactions taken into account
 and, as intermediate conclusion, of building up the final spectrum of the 
 partons produced in the hard processes.\par
 A former version of this investigation was presented at a Workshop in Turin,[1]
 in order to save space we shall make systematic reference to this paper, that 
 will be called T, where also a larger bibliography is listed.

  \vskip .5pc
  \section {Transport equation for the produced partons}
  \vskip .5pc
  
 The guiding idea can be stated in this way: the incoming hadron has a
 distribution of partons that interact with the partons of the heavy nucleus,
 the produced partons may be also partially reabsorbed, so we write a transport
 equation for the population of the secondary partons that go {\it forward i.e.}
 in the direction of the hadron.
  We introduce a parameter 
  $\tau$: the depth at which the nucleon has drilled
  the nucleus, or the mean number of hit partons of the nucleus. 
  The parameters $u_i$ are the relevant coordinated of the primary partons,
  usually impact parameter and rapidity, $u=({\bf b},y)$ the parameters $v_i,w$ 
  are the relevant coordinates of the secondary partons they may be either
  again impact parameter and rapidity or transverse momentum and rapidity 
  $v=({\bf k},y)$.
  As it was done in a simpler form in T a transport equation is written:
  \begin {eqnarray} 
&P_r^n&(\{u\};v_1,\ldots{,v_r;\tau+\Delta\tau})
=P_r^n(\{u\};v_1,\ldots{,v_r;\tau})
\cr\cr
&+&\sum_{i=1}^n\sum_{s=1}^r
P_{r-1}^n(\{u\};v_1,\ldots{,v_{s-1},v_{s+1},}\ldots{,v_r;\tau})
E(u_i;v_s;\tau)\Delta\tau
\cr\cr
&+&\sum_{i=1}^{n}\int{dw}P_{r+1}^n(\{u\};v_1,\ldots{v_r,;w
;\tau})\,A(u_i;w;\tau)\Delta{\tau}
\cr\cr
&-&\sum_{i=1}^n\sum_{s=1}^r P_r^n(\{u\};v_1,\ldots{,v_r;\tau})
A(u_i;v_s;\tau)\Delta{\tau}
\cr\cr
&-&\sum_{i=1}^n\int{dw}P_r^n(\{u\};v_1,\ldots{,v_r;\tau})
E(u_i;w;\tau)\Delta\tau\crcr
&\quad&\quad\quad\quad \{u\}=u_1\dots u_n
 \end {eqnarray}
 
  $E\,,\,A$ are emission and absorption probabilities, which may
   depend on $\tau$.
 Only one
  emission or absorption is involved, four basic steps are foreseen\par
  $ r-1 \to r\quad \bullet\quad r+1 \to r \quad\bullet\quad
   r\to r-1\quad\bullet\quad r\to r+1$\par
   (The overall impact parameter $\beta$, is fixed and it will not be 
   written).
   The equations solved by defining a generating functional for every given set
   of primary coordinates $\{u\}$:
   \begin {equation}
 {\cal F}^n[\{u\};I;\tau] =\sum_r {1\over{r!}}\int dv_1,\dots,dv_r
  I(v_1)\dots 
  I(v_r) P_r^n(\{u\};v_1,\dots ,v_r;\tau)\; 
  \end {equation}
  and an equation in the continuum limit $\Delta\tau\to 0$.
 
    \begin {eqnarray} 
 {\partial\over{\partial \tau}}{\cal F}^n[\{u\};I;\tau]&=&
 \sum_{i=1}^n{\cal F}^n[\{u\};I;\tau]\int
 dw I(w) E(u_i,w;\tau)\cr\cr
 &+&\sum_{i=1}^n\int dw  A(u_i,w;\tau) {\delta\over{\delta I(w)}}
 {\cal F}^n[\{u\}I;\tau]\cr\cr
 &-&\sum_{i=1}^n{\cal F}^n[\{u\};I;\tau]\int dw E(u_i,w;\tau)\cr\cr
   &-&\sum_{i=1}^n\int dw  A(u_i,w;\tau)I(w) {\delta\over{\delta I(w)}}
    {\cal F}^n[\{u\};I;\tau] \quad
     \end {eqnarray} 
     The solution for ${\cal F}$ with the initial condition that there are no
     secondaries at all for $\tau=0$ yields the expression for $P$ which
     corresponds to a Poisson distribution, in condensed form 
     $$P_r=p_1\dots p_r \e^{-\int pdv}$$
    where
   \begin {equation}
 p^{(n)}(\{u\};v;\tau_f)=\sum_{i=1}^{n}\int_0^{\tau_f}
dt\,E(u_i;v;t)
\exp\left[-\sum_{i=1}^n\int_t^{\tau_f}\,dt'\,A(u_i;v;t')\right].
\end {equation}
The expression of $p^{(n)}$ is the one-body distribution of secondaries at fixed
distribution of the primaries, is not yet an observable result, one must in fact
sum over the different
partonic structure of the colliding hadron. If one assumes, as
it is sometimes done, a Poissonian distribution for the primary partons:
\begin {eqnarray}
 G_{1\cdots n}&=&C(u_1)\dots C(u_n)
 \exp\Bigl [-\int C(u)du\Bigr ]\,,\cr\cr
 &\sum_n& {1\over {n!}}\int G_{1\cdots n}d^n u=1.
\end {eqnarray}
one gets an explicit although complicated expression per the distribution of
the secondaries; in particular the one-body distribution is

\begin{eqnarray}
&D_1&(v;\tau_f)\cr\cr 
&=&\sum_{n=0}^{\infty}{1\over{n!}}\exp\Bigl[-\int{du\,C(u)}\Bigr]
\int{du_1\cdots{du_n}\,C(u_1)\cdots{C(u_n)}}
\,p^{(n)}(\{u\};v;\tau_f)\cr\cr
&=&
\int_0^{\tau_f}dt\,\int du\,C(u)\,E(u;v;t)\,
\exp{\left[-\int_t^{\tau_f}dt'A(u;v;t') \right]}\cr\cr
&\times &\exp\left\{-\int{du\,C(u)}\,\left(1-\exp{\left[-\int_t^{\tau_f}dt'
A(u;v;t')
\right]}\right)\right\}.
\end{eqnarray}
By calculating the 2-body function $D_2(v_1,v_2)$, it is found that the
distribution is no longer Poissonian even at fixed $\beta$; in fact
$$ {\cal C}_2(v_1,v_2)\equiv D_2(v_1,v_2)-D_1(v_1)D_1(v_2)\neq 0$$
The choice of the Poissonian distribution for
the primary partons is only a simplified example, one could deal with much more
general distributions by using the formulation in term of generating 
functionals [2].

  \vskip .5pc
  \section {Emission and absorption coefficients}
  \vskip .5pc
  
  When only the elastic scattering among partons was taken as the elementary
  dynamical process the basic ingredient was the elastic cross section, now it
  is necessary to feed into the formalism the elementary $ 2\rightleftharpoons
  3$ processes.\par
  If we want this process to be described in perturbative terms there are
  kinematical limitations (in particular the sub-energy of every pair of parton
  must be large) so that the process may be described in term of amplitude 
  $M_{gg\to ggg}$ of the the nonlocal Lipatov vertex[3], its absolute square may 
be
  brought into the form:
  \begin{equation}
  \big|M_{gg\to ggg}\big|^2 ={54 g^6}
 {{\hat s^2}\over{k^2_{0\bot}k^2_{1\bot}k^2_{2\bot}}}\quad
 \sum_i {\bf k}_{i\bot}={\bf p}_a+{\bf p}_b=0
 \end {equation}
  Now the rest of the calculation may be only indicated: since we work, see T,
  in the impact parameter space the Fourier transform of $M_{gg\to ggg}$ with
  respect to the momentum transfer between the primary partons must be
  calculated, with this procedure the coefficient $E$ is obtained, then pure
  kinematics must yield the absorption coefficient $A$.\par
  The spectrum of the produced particles is then obtained, this result can be
  also read as a modification of the primaries' population induced by the hard 
  scattering among partons.
  
   \vskip .5pc
\section*{Acknowledgments}
 \vskip .5pc
This work has been partially supported by the Italian Ministry of University
 and of Scientific and Technological Research by means of the {\it Fondi per la
 Ricerca scientifica - Universit\`a di Trieste }.

\vfill
\eject  
\end{document}